\documentclass[10pt,letterpaper]{article}
\usepackage{opex3}
\usepackage{graphicx}
\usepackage{dcolumn}
\usepackage{bm}
\usepackage{cite}

\begin{document}




\newcommand \beq {\begin{equation}}
\newcommand \eeq {\end{equation}}
\newcommand \beqa {\begin{eqnarray}}
\newcommand \eeqa {\end{eqnarray}}
\newcommand \bE {{\bf E}}
\newcommand \bc {{\hat{c}}}

\newcommand \Soned {S_{11}^{\rm (1d)}}
\newcommand {\btdnote}[1]{{\bf\color{red}[btd:~#1]}}
\newcommand {\pjfnote}[1]{{\bf\color{blue}[pjf:~#1]}}
\newcommand \jatmossci {{J.\ Atmosph.\ Sci.}}
\newcommand \jgr {{J.\ Geophys.\ Res.}}
\newcommand \jqsrt {{J.\ Quant.\ Spec.\ Rad.\ Trans.}}
\newcommand \optlett {{Opt.\ Letters~}}
\newcommand \optex {{Opt.\ Express~}}

\newcommand{\bfr}{{\mathbf{r}}}
\newcommand{\ee}{{\mathrm{e}}}
\newcommand{\calG}{{\mathcal{G}}}
\newcommand{\dd}{{\mathrm{d}}}
\newcommand{\PsiGML}{{\Psi_{\mathrm{GML}}}}
\newcommand{\Heff}{{H_{\mathrm{eff}}}}
\newcommand{\barE}{\overline{E}}
\newcommand{\barP}{\overline{P}}
\newcommand{\barphi}{\overline{\varphi}}

\newcommand{\newtext}[1]{{\color{red}#1}}

\title{Light scattering by hexagonal columns in the discrete dipole approximation \\ }

\author{P. J. Flatau$^{1, *}$ and B. T. Draine$^2$}
\address{$^1$University of California San Diego, Scripps Institution of Oceanography, \\ La Jolla, CA, 92093 USA,  $^2$Princeton University Observatory, Princeton, \\ New Jersey 08544-1001, USA}
\email{$^*$pflatau@ucsd.edu}




\begin{abstract}
Scattering by infinite hexagonal ice prisms is calculated using Maxwell's
equations in the discrete dipole approximation for size parameters 
$x=\pi D/\lambda$ up to $x=400$  ($D=$ prism diameter).
Birefringence is included in the calculations.
Applicability of the geometric optics approximation
is investigated. Excellent agreement between wave optics
and geometric optics is observed for large size parameter in the outer part of the 22 degree halo feature. For smaller ice crystals halo broadening is predicted, and there is appreciable ``spillover'' of the halo into shadow  scattering angles $<22$ degrees.
Ways to retrieve ice crystal sizes are suggested based on the full width at half-maximum of the halo, the power at  $<22$deg,
and the halo polarization.
\end{abstract}



\bibliographystyle{osajnl}
\bibliography{bibhex}


\section{Introduction}

\subsection{Ice crystals}

Microphysical properties of cirrus clouds 
\cite{lynch2001cirrus,cotton2010storm}, 
such
as the scattering asymmetry parameter and 
single scattering albedo of ice crystals,
affect climate feedback processes
\cite{stephens1990relevance}.
These microphysical properties are governed by ice crystal shape, size,
refractive index, and the mass of ice crystals per unit volume of air.
Simple ice crystals are
non-spherical, often hexagonal columns or plates, sometimes with
irregularities such as inclusions or voids 
\cite{macke1996influence,shcherbakov2013why}. 
Scattering by more complicated crystal shapes can also be calculated
\cite{bi2014accurate}.
Besides their importance for atmospheric radiative transfer, they are
also responsible for intricate optical displays
\cite{greenler1980rainbows,tape1994atmospheric,tape2006atmospheric}
and polarization effects  \cite{konnen1985polarized}.
Their single scattering properties play an important role
in remote sensing in visible and infrared light as well as in
microwaves 
\cite{evans1995microwave}.
Light scattering properties of
ice crystals have been typically
calculated in the geometric optics 
(GO) approximation by ray-tracing algorithms 
\cite{tape1994atmospheric,macke1993scattering,macke1996single,takano1989solar,cai1982polarized,wendling1979scattering,borovoi2007light}.
When crystal dimensions are comparable to the incident wavelength $\lambda$, 
diffractive effects have been
estimated
using circular cylinders 
\cite{mishchenko1999big,lee2003use}
and combinations of several techniques
\cite{yang2013spectrally,bi2014accurate}.


\begin{figure}[t]
\begin{center}
                
\includegraphics[width=7.5cm,angle=0,clip=true,trim=0.0cm 0.0cm 0.0cm 0.0cm]{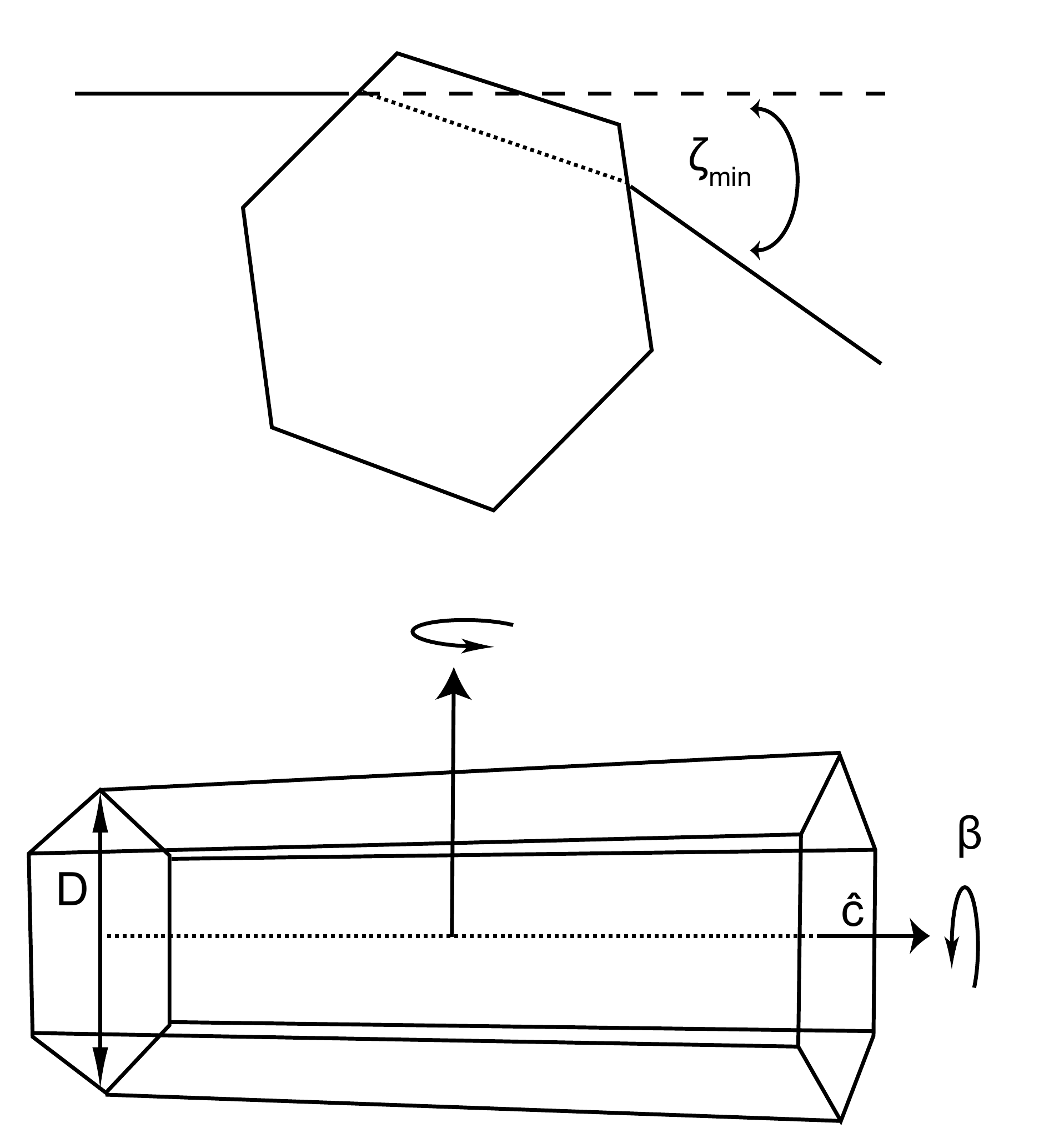}
\caption{\label{fig:hex} Hexagonal ice crystal with vertex-to-vertex
  diameter D. There are six side faces and two end (basal) faces.  The
  {\it c}-axis is parallel to the sides and perpendicular to
  the end faces. Prisms are rotated randomly around the {\it c}-axis.  In the
  GO approximation, halos discussed in this paper peak at the
  minimum deviation angle $\zeta_{\rm min}$ for which once-refracted
  light is propagating parallel to a side face before the second
  refraction.}
\end{center}
\end{figure}
Analytic
solutions to Maxwell's equations are known only for special
geometries such as spheres, spheroids, or infinite cylinders, so
approximate methods are in general required.
The Discrete Dipole Approximation (DDA) is one such 
method.
The basic theory of the DDA has been presented elsewhere
\cite{flatau1994discrete}.
Conceptually, the DDA consists of approximating the target of interest
by an array of polarizable points.
Once the polarizability tensors are specified, Maxwell's
equations can be solved 
to arbitrary accuracy for the array, and the scattering problem of interest
can be solved to arbitrary accuracy by reducing the dipole separation
$d\rightarrow0$ (within computational limits).
Developed originally to study scattering from isolated, finite
structures such as dust grains, the DDA was extended to treat singly
or doubly periodic structures
\cite{Chaumet+Rahmani+Bryant_2003,draine2008discrete}; 
we also
generalized the scattering amplitude matrix and the 4$\times$4 Mueller matrix
to describe scattering by such periodic targets.
The accuracy of DDA calculations using the open-source code DDSCAT was
demonstrated by comparison with exact results for infinite cylinders
and infinite slabs.
A fast method for using the DDA solution to obtain fields within and near
the target was also presented
\cite{flatau2012fast}.

In this paper we 
use
the DDA for periodic
targets to calculate light scattering
properties of infinite hexagonal ice crystals with size parameter
$x\equiv\pi D/\lambda\leq400$, where $D$ is the diameter.
Such infinite hexagonal prisms approximate light scattering by 
atmospheric ice crystals with a large aspect ratio
\cite{cai1982polarized}.
We show that finite diameter ice crystals have scattering properties
that deviate systematically from geometric 
optics; these deviations 
can be used to infer the cystal diameter $D$ from measured halos. 

\subsection{Geometry and relation to atmospheric optics}

Water ice crystals form hexagonal columns, with the column
symmetry axis parallel to the crystalline {\it c}-axis.
Long ice crystals fall with their {\it c}-axis 
oriented horizontally
\cite{greenler1980rainbows,tape1994atmospheric}.
Such ice crystals
will be randomly rotated about the {\it c}-axis and about a vertical axis
perpendicular to the {\it c}-axis 
\cite{tape1994atmospheric}.
In our calculations we include the effects of random rotations around
the {\it c-}axis, but we consider only two values of the
angle $\Theta$ between the incident beam and the {\it c-}axis.
Real atmospheric halos also include a contribution from reflections
by hexagonal prism end faces.
End faces are not included in the present DDA calculations because we
assume infinitely long crystals to allow
use of the periodic boundary condition version of
the DDA to study very large values of $D$.

We consider an infinitely long hexagonal prism, with light incident at
an angle $\Theta$ relative to the {\it c}-axis.  If the sides are
numbered $j=1,...,6$ in order, then rays entering through side $j$ and
exiting through side $j+2$ will be refracted, with the emergent ray
scattered by an angle $\theta_s$.  
We will refer to the scattering associated with light passing
through sides $j$ and $j+2$ (this is path 35 in
\cite{tape1994atmospheric})
as the ``$22^\circ$ feature''.
(The term ``$22^\circ$ halo'' is a standard term for the halo which
is generated by light passing through sides $j$ and $j+2$ of 
randomly-oriented crystals.
However, here we deal with 
preferentially oriented crystals.)
The photon momentum parallel to the
{\it c}-axis is unchanged.  As the hexagonal prism is rotated around
the {\it c}-axis, the minimum scattering angle $\theta_{\rm s,min}$
occurs when the ray passes through the prism traveling parallel to one
of the sides; for randomly-rotated prisms, the halo peak is at
$\theta_{\rm s,min}$.  Let $\zeta$ be the deflection of the ray
projected on the basal plane.
The minima of $\zeta$ and $\theta_s$ are 
\cite{Mariotte1681,adam2011mathematics}
\beqa \label{eq:zeta_min}
\zeta_{\rm min}&=& 
2\arcsin\left[\frac{(n^2-\cos^2\Theta)^{1/2}}{2\sin\Theta}\right]
-60^\circ
\\ \label{eq:thetas_min}
\theta_{\rm s,min}&=&\arccos\left[1+\sin^2\Theta(\cos\zeta_{\rm min}-1)\right]
~~~,
\eeqa
where $n$ is the real part of the refractive index.
At $\lambda=550$nm and $T=-25$C, H$_2$O ice has 
$n_o(\bE\perp \bc)=1.3112$, $n_e(\bE\parallel \bc)=1.3126$
for ``ordinary'' and ``extraordinary'' polarizations
\cite{Ehringhaus_1917,konnen1985polarized}.
For $\Theta=90^\circ$,
the halo peak is at
$\theta_{\rm s,min}=21.930^\circ$ and $22.037^\circ$ for the ordinary
and extraordinary polarizations, respectively. 

\section{Discrete dipole approximation method}
\subsection{Scattering}
The scattering properties of 
targets with 1-dimensional translational symmetry
are described by the generalized Mueller matrix elements
$S_{ij}^{\rm (1d)}$ 
relating the Stokes parameters of the scattered wave to the Stokes
parameters of the incident plane wave
\cite{draine2008discrete}.
For unpolarized incident light,
Stokes parameters $I$ and $Q$ for the scattered wave are proportional
to $\Soned$ and $S_{21}^{\rm(1d)}$, respectively, and the polarization 
$P=Q/I=S_{21}^{\rm(1d)}/\Soned$.

The 
$S_{ij}^{\rm (1d)}$
are directly related to differential scattering cross sections
per unit length.
In particular,
\beq
\Soned = \frac{k_0}{\sin\Theta}
\frac{d^2C_{\rm sca}}{d\zeta dL}
~~~,
\eeq
where 
$C_{\rm sca}$ is the scattering cross section for unpolarized light,
$k_0\equiv2\pi/\lambda$, and
$0\leq\zeta<2\pi$ is the azimuthal angle around the
``scattering cone'', with $\zeta=0$ for forward scattering.
The scattering angle $\theta_s\geq0$
is related to $\zeta$ by
\beq \label{eq:theta_s}
\cos\theta_s = 1 - \sin^2\Theta(1-\cos\zeta)
~~~;
\eeq
for given $\Theta$, the largest allowed scattering angle is
$\theta_{\rm s,max}=\arccos(1-2\sin^2\Theta)$.

For each allowed $\theta_s$, there are two values of $\zeta$
satisfying (\ref{eq:theta_s}): let these be $0\leq\zeta_1\leq\pi$ and
$\zeta_2=2\pi-\zeta_1$.
The usual differential scattering cross section is related
to $\Soned$ by
\beqa \nonumber
\frac{d^2C_{\rm sca}}{d\theta_s dL} &=&
\frac{\sin\theta_s}{k_0\sin\Theta\sin\zeta_1}
\left[\Soned(\zeta_1)+\Soned(\zeta_2)\right]
\\ \label{eq:dCsca/dtheta_from_S11}
&=&
\frac{\sin\theta_s\sin\Theta\left[\Soned(\zeta_1)+\Soned(\zeta_2)\right]}
     {k_0\sqrt{(1-\cos\theta_s)(2\sin^2\Theta+\cos\theta_s-1)}}
~~.~~
\eeqa
After averaging over target rotations around the {\it c}-axis,
the present problem has reflection symmetry, 
$\Soned(\zeta_2)=\Soned(\zeta_1)$.
For forward scattering ($\zeta=\theta_s=0$) 
Eq.\ (\ref{eq:dCsca/dtheta_from_S11})
becomes
\beq
\frac{d^2C_{\rm sca}}{d\theta_s dL}(\theta_s=0) =
\frac{2}{k_0}\Soned(\zeta=0)
~~~,
\eeq
and for $\zeta=\pi$
\beq
\frac{d^2C_{\rm sca}}{d\theta_s dL}(\theta_s=\theta_{\rm s,max})
=
\frac{2}{k_0}\Soned(\zeta=\pi)
~~~.
\eeq

\subsection{Accuracy}


\begin{table*}[b]
\centering
\caption{\label{tab:DDA_params} Parameters for DDA calculations}
\footnotesize
\begin{tabular}{c c c c c c c c}
$x$ & $D (\mu{\rm m})^{a}$ & $N^{b}$ & $d (\mu{\rm m})^{c}$ & $|m_e|k_0d$ &
$tol\,^{d}$ &$N_{\rm iter}\,^{e}$ &CPU hrs\\
\hline
25  & \ 4.38             & \ \ 11903  & 0.03232 & 0.485 & $10^{-4}$ & \ \ 128 & \ \ \ \ \ \ \ 0.085 \\ 
50  & \ 8.75             & \ \ 47488  & 0.03237 & 0.485 & $10^{-4}$ & \ \ 837 & \ \ \ \ \ \ \ 0.748 \\
100 & 17.5\ \ \,             & \ 189724 & 0.03239 & 0.486 & $10^{-4}$ & \ 4196 & \ \ \ 25.8\ \ \\
150 & 26.3\ \ \,              & \ 426708 & 0.03240 & 0.486 & $10^{-4}$ & 10890& \ 114.\ \ \ \\
200 & 35.0\ \ \,            
                       & \ 758436 & 0.03240 & 0.486 & $10^{-4}$ & 16887 & \ 384.\ \ \ \\
    &                  & 1663744& 0.02188 & 0.328 & $5\times10^{-5}$&21178& \ 958.\ \ \ \\
300 & 52.5\ \ \,             & 1663744& 0.03282 & 0.492 & $10^{-4}$& 37878& 1393.\ \ \ \\
400 & 70.0\ \ \,             & 3031748& 0.03241 & 0.486 & $5\times10^{-4}$& 12520 & 3301.\ \ \ \\
\hline
\multicolumn{4}{l}{$^a$~ $D=$ vertex-vertex diameter}&
\multicolumn{4}{l}{$^d$~ DDSCAT error tolerance}\\
\multicolumn{4}{l}{$^b$~ $N=$ number of dipoles in one target unit cell}&
\multicolumn{4}{l}{$^e$~ avg.\ number of iterations per orientation}\\
\multicolumn{4}{l}{$^c$~ $d=$ interdipole spacing}&
\multicolumn{4}{l}{~~~~and incident polarization}\\
\end{tabular}
\end{table*}

We have performed calculations using DDSCAT
\cite{flatau1994discrete,draine2008discrete} 
for regular infinite hexagonal prisms
with size parameters
up to $x=\pi D/\lambda=400$.
Birefringence is fully included in the DDA treatment.  We use
complex refractive index
\beqa
m_o(\bE\perp \bc)&=&1.3112+10^{-5}i
\\
m_e(\bE\parallel \bc)&=&1.3126+10^{-5}i ~~~.
\eeqa
The weak absorption [Im$(m)=10^{-5}$], introduced to
accelerate the conjugate gradient solution, 
is small enough that
the scattering results are not affected appreciably.

The DDA calculations were done for 10 different rotations
of the target around the {\it c}-axis, at intervals of
$\Delta\beta=3^\circ$; with the symmetry of the target, these 10 orientations
are used to average over rotations of the target
around the {\it c}-axis.

Calculations for larger size parameters required extensive computer resources and several days of computer time. We used the OpenMP option in DDSCAT.
Parameters for the dipole arrays are given in Table \ref{tab:DDA_params},
including the dipole spacing $d$, and 
the number $N$ of dipoles in a single hexagonal layer (the
``target unit cell'').

We use DDSCAT with conjugate gradient algorithm
 ``GPBICG'' 
\cite{Tang+Shen+Zheng+Qiu_2004,Chaumet+Rahmani_2009}
to iteratively solve the system of linear equations.
Iteration terminates
when a specified error tolerance {\it tol} is achieved; {\it tol}
is listed in Table \ref{tab:DDA_params}, as well as $N_{\rm iter}$,
the average number of iterations required to reach the specified error
tolerance for a single orientation and incident polarization.
For a given $x$, 
the required CPU time is approximately proportional to $N_{\rm iter}\times N\ln(N)$.
For the largest problem, $x=400$, with
$N>3\times10^6$ dipoles, the error tolerance was raised to
$tol=5\times10^{-4}$ in order to lower $N_{\rm iter}$ and reduce the required
CPU time.
Table \ref{tab:DDA_params} also gives
the total CPU time required for convergence (for 10 orientations and two polarizations)
for single-precision 
calculations using 4 cores running at 2.6 GHz.


\begin{figure}[t]
\centering
\includegraphics[width=8.5cm,angle=0,%
                 clip=true,trim=0.5cm 5.0cm 0.5cm 0.5cm]
                {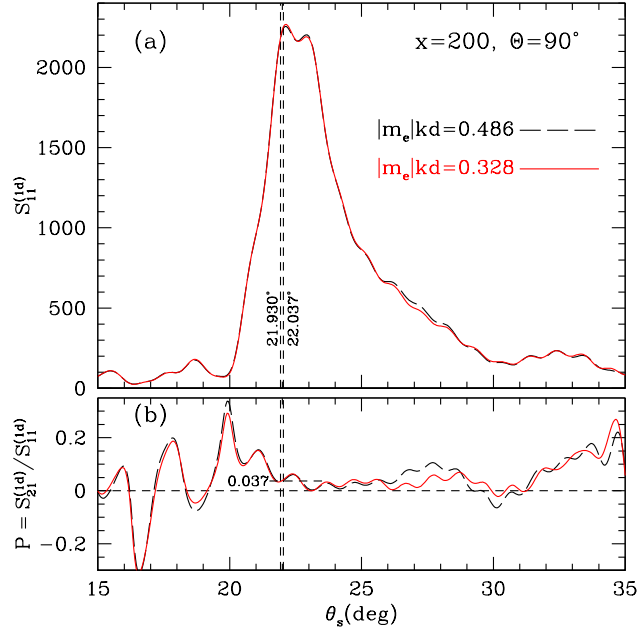}
\caption{\label{fig:accuracy}
         Discrete dipole approximation results for incidence angle $\Theta=90^\circ$
         and $x=200$, averaged over
         rotations around the
         {\it c}-axis (sampled at $\Delta\beta=3^\circ$ intervals),
         for 2 values of
         $d$ (see Table \ref{tab:DDA_params}).
         Vertical dashed lines show the geometric optics halo angles
         $\theta_{\rm s,min}$ for ordinary and extraordinary
         polarizations.
         (a) Muller matrix element $S_{11}^{(1d)}$ 
         (total halo intensity $I\propto S_{11}^{(1d)}$)
         (b) Fractional polarization $P$.
         The short dashed line indicates
         the polarization ($P=0.037$) of the $22^\circ$ halo peak in the GO limit.
         }
\end{figure}
\begin{figure*}
\centering
\includegraphics[width=9.5cm,angle=270,
                 clip=true,trim=0.5cm 0.5cm 0.5cm 0.4cm]
                {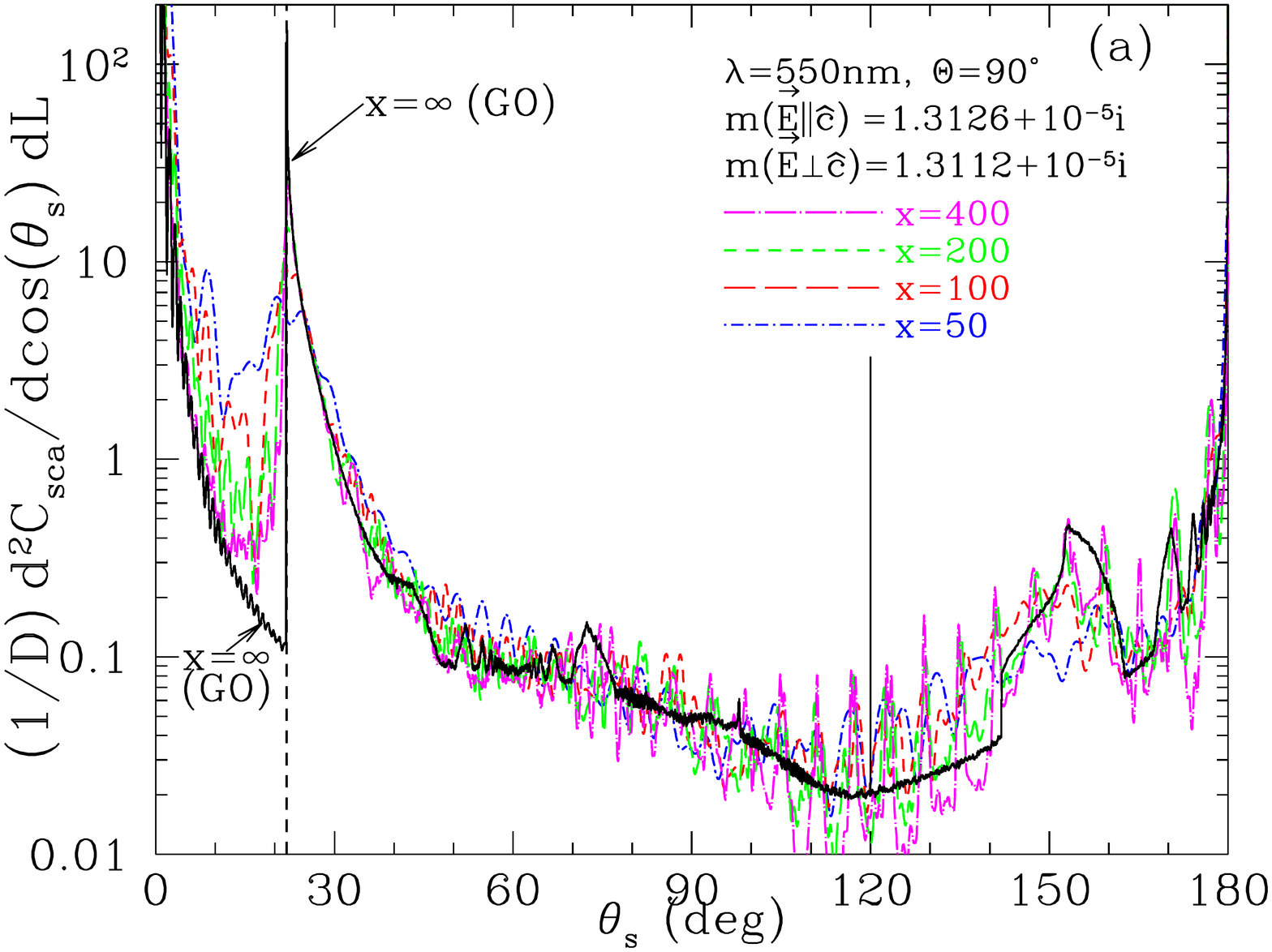}
\includegraphics[width=9.5cm,angle=270,
                 clip=true,trim=0.5cm 0.5cm 0.5cm 0.4cm]
                {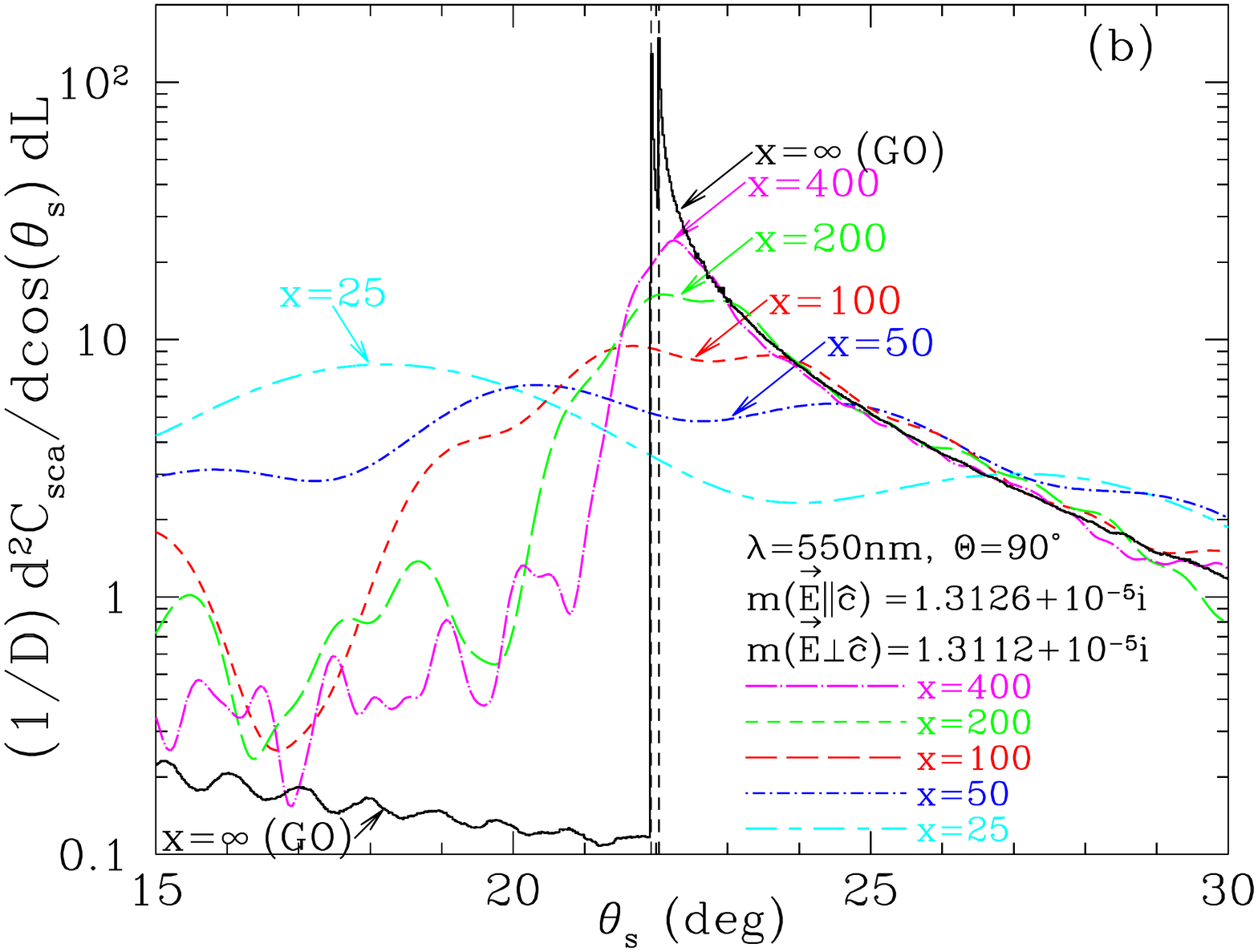}
\caption{\label{fig:S_11_Theta=90}
  $(1/D)d^2C_{\rm sca}/d\cos\theta_sdL$ as a function of
  scattering angle $\theta_s$ for hexagonal prism with light incoming
  perpendicular to {\it c}-axis, averaged over prism rotations
  around the {\it c}-axis.  
  (a) Full range of $\theta_s$; 
  (b) zoom on $15^\circ<\theta_s < 30^\circ$.
  DDA results are shown for selected
  values of $x\equiv\pi D/\lambda$, as well as GO results (see text).
  Vertical dashed lines show the expected position $\theta_{\rm s,min}$
  of the inner edge
  of the $22^\circ$ halo for ordinary and extraordinary polarizations
  in the ray-tracing limit.
  }
\end{figure*}

Previous studies comparing
DDA results with exact solutions for spheres 
\cite{flatau1994discrete}
indicate
that DDA results will be accurate provided 
$|m|k_0d<0.5$, 
with the DDA results converging
to the exact solution as $d\rightarrow0$.  The 
$|m|k_0d<0.5$
condition is satisfied in the calculations presented here,
so that the DDA
should provide an accurate representation of the infinite prism.

As a check, we have repeated the calculations for $x=200$ for
smaller
interdipole spacing $d$.
Figure~\ref{fig:accuracy} shows $S_{11}$ and $S_{21}$ for
$15^\circ<\theta_s<35^\circ$ for DDA realizations
with $|m_e|k_0d=0.486$ and $0.328$;
the two calculations are in excellent agreement.
For a given target, we expect the DDA results (if fully converged)
to be exact in the
limit $d\rightarrow0$, with the error varying approximately in
proportion to $d$.
Better accuracy can be obtained by reducing $d$, but this becomes
computationally expensive for large $x$.


\section{Results}
\subsection{Halo features}
Figure~\ref{fig:S_11_Theta=90} shows
$d^2C_{\rm sca}/d\cos\theta_sdL$ calculated for $\Theta=90^\circ$,
for selected values of the scattering parameter $x\equiv\pi D/\lambda$.
In the GO calculation, the inner edge of the $22^\circ$ halo
is at $21.93^\circ$ and $22.04^\circ$ for ordinary and extraordinary
polarizations.
In the wave optics calculation for $x=25$, the peak occurs at $\sim$$18^\circ$,
well interior to the inner edge of the GO halo.
As $x$ is increased, the halo power becomes more concentrated, and closer
to the expected halo angle $22^\circ$.
For $x=50$ we find a double-peaked halo, centered
near $22^\circ$.
For $x\geq100$ the peak becomes increasingly
narrow and pronounced as $x$ is increased.

The peaks occuring every $6^\circ$ from $33^\circ$
to $177^\circ$
in Fig.~\ref{fig:S_11_Theta=90}(a)
are produced by specular reflection.
Each orientation 
$\beta$
contributes specular reflection peaks at specific
scattering angles.
Because we
sample rotations around the {\it c}-axis at $\Delta\beta=3^\circ$ intervals,
the peaks from different orientations are 
separated by $\Delta\zeta=2\Delta\beta=6^\circ$ 
(for $\Theta=90^\circ$, $\Delta\theta_s=\Delta\zeta$).
These peaks, with diffractive broadening
$(\delta\theta_s)_{\rm diff}\approx \lambda/D=180^\circ/x$,
become narrower as $x$ is increased, and hence are most
conspicuous for the $x=400$ case.
For finer rotational sampling $\Delta\beta\rightarrow0$, 
these specular reflection
peaks would blend into a continuum.

Also shown in Fig.~\ref{fig:S_11_Theta=90} is $\Soned$ obtained from
a GO calculation using a ray-tracing code written by A.\ Macke
\cite{macke1996single},
with the scattering binned
with $\Delta\theta_s=0.02^\circ$.
The conspicuous spike in the GO results at $\theta_s=120^\circ$ comes
from rays that enter through side $j$, undergo internal reflections
from sides $j+1$, $j+2$, $j+4$, and $j+1$, and exit through
side $j+5$ with a deflection of exactly $120^\circ$.
With only $\sim$0.1\% of the scattered 
power in this feature,
diffractive broadening washes out the corresponding scattering
in the DDA results.
%
Overall, the DDA results appear to be
clearly converging to the GO limit as
$x\rightarrow\infty$.

\begin{figure*}[htbp]\centering
\includegraphics[width=9.5cm,angle=270,
                 clip=true,trim=0.5cm 0.5cm 0.5cm 0.4cm]
                {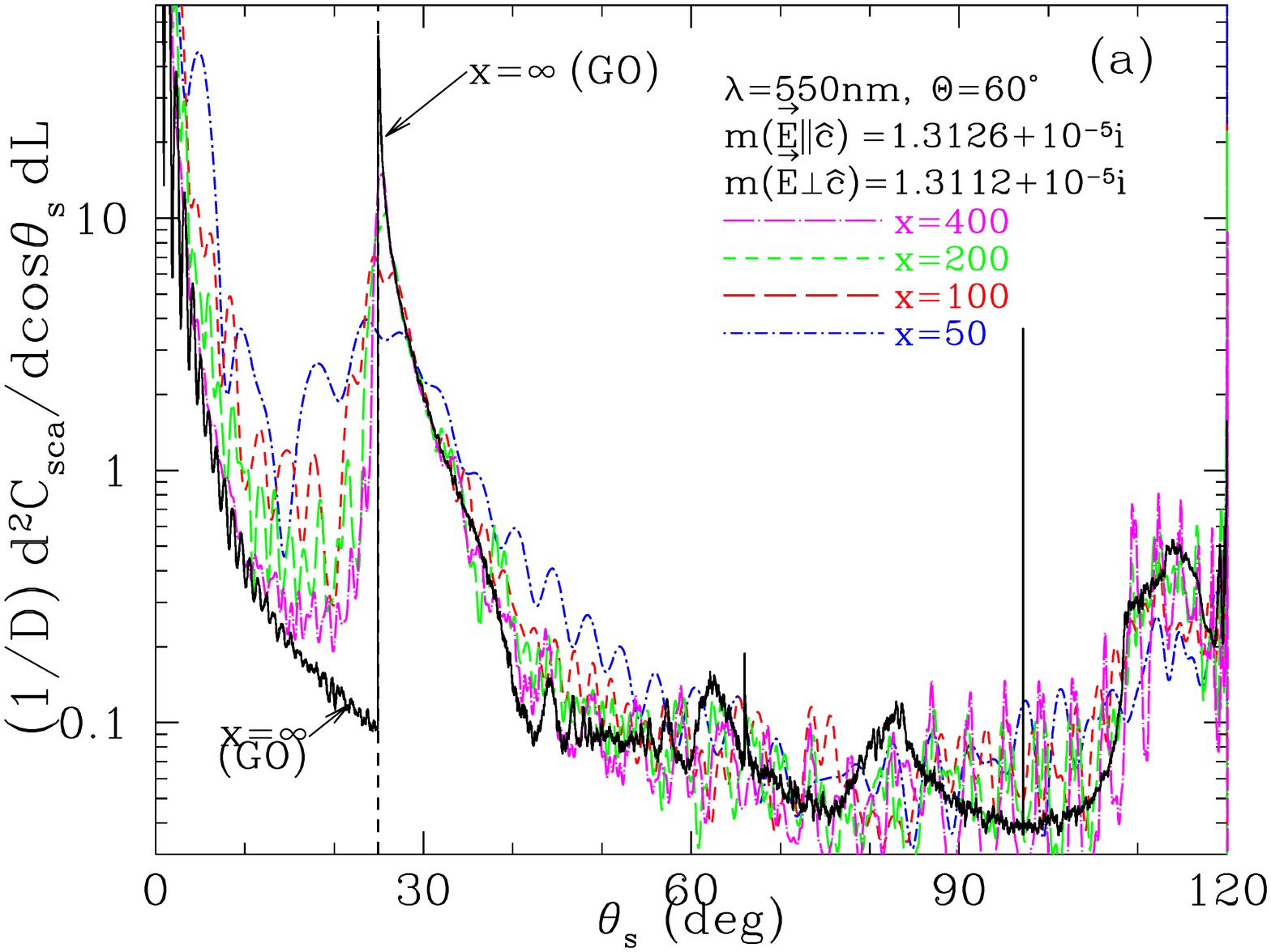}
\includegraphics[width=9.5cm,angle=270,
                 clip=true,trim=0.5cm 0.5cm 0.5cm 0.4cm]
                {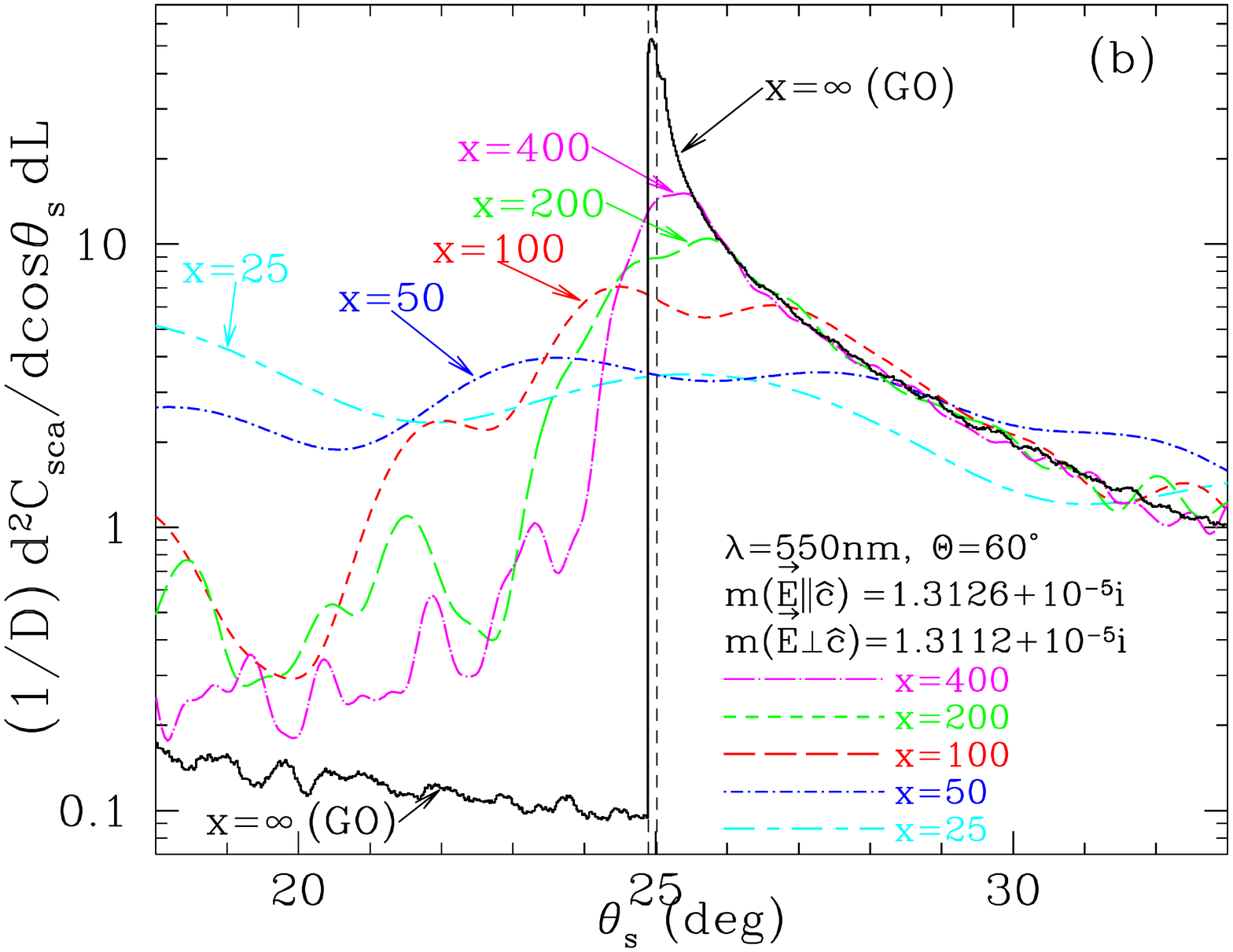}
\caption{\label{fig:S_11_Theta=60} 
  Similar to Fig.~\ref{fig:S_11_Theta=90}, but for
  oblique incidence $\Theta=60^\circ$.
  (a) Full range of allowed scattering angles $\theta_s$ ($0-120^\circ$);
  (b) zoom on $18^\circ<\theta_s<33^\circ$.
  For $\Theta=60^\circ$ the 
  GO cusps are at $24.896^\circ$ and $25.017^\circ$ for
  ordinary and extraordinary rays.
  Note the good agreement between GO and the $x=400$ results
  for $\theta_s>26^\circ$.
  }
\end{figure*}

\begin{figure}[htbp]\centering
\includegraphics[angle=270,width=12.5cm,
                 clip=true,trim=0.5cm 0.5cm 0.5cm 0.5cm]
                {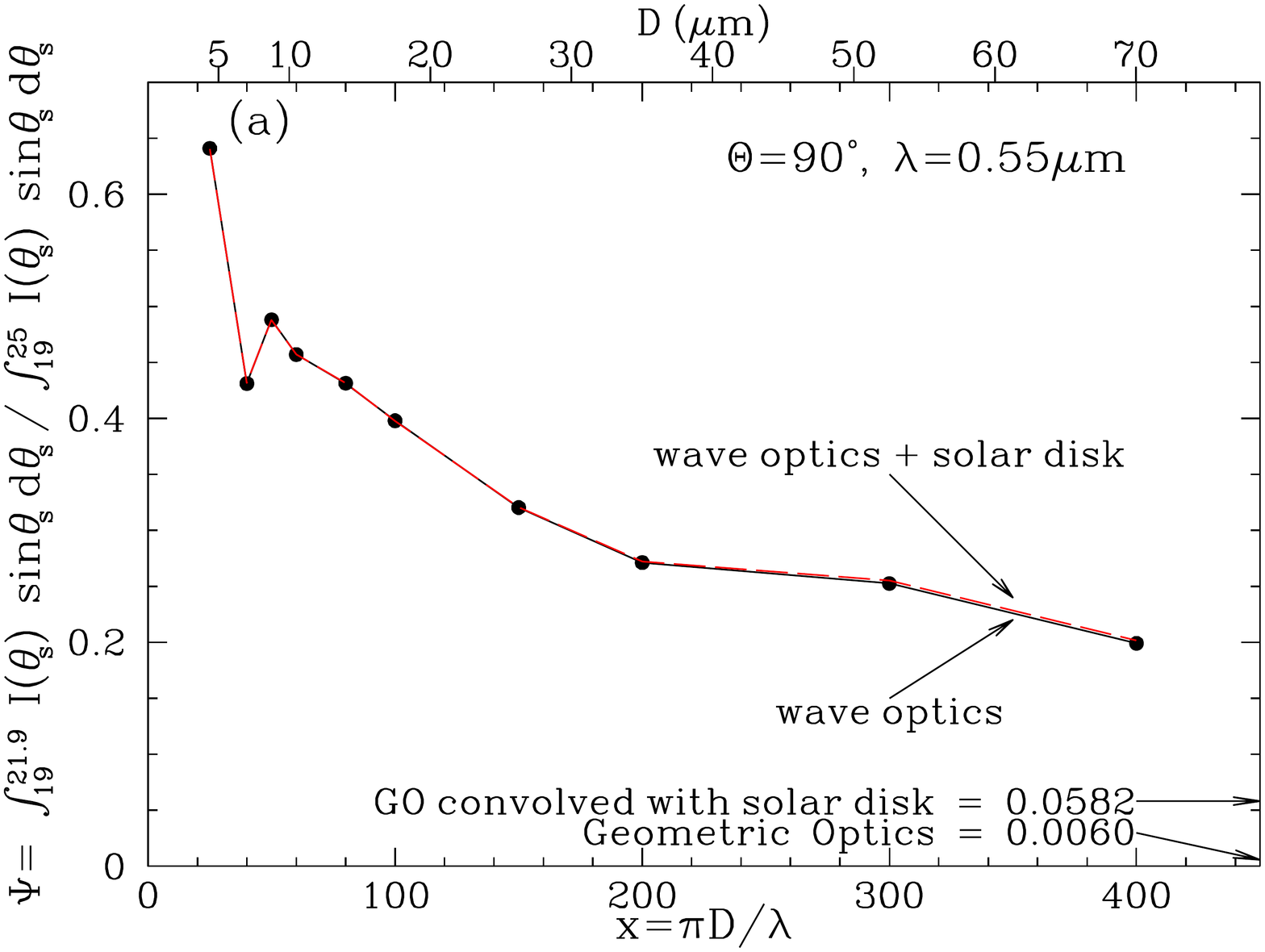}
\includegraphics[angle=270,width=12.5cm,
                 clip=true,trim=0.5cm 0.5cm 0.5cm 0.5cm]
                {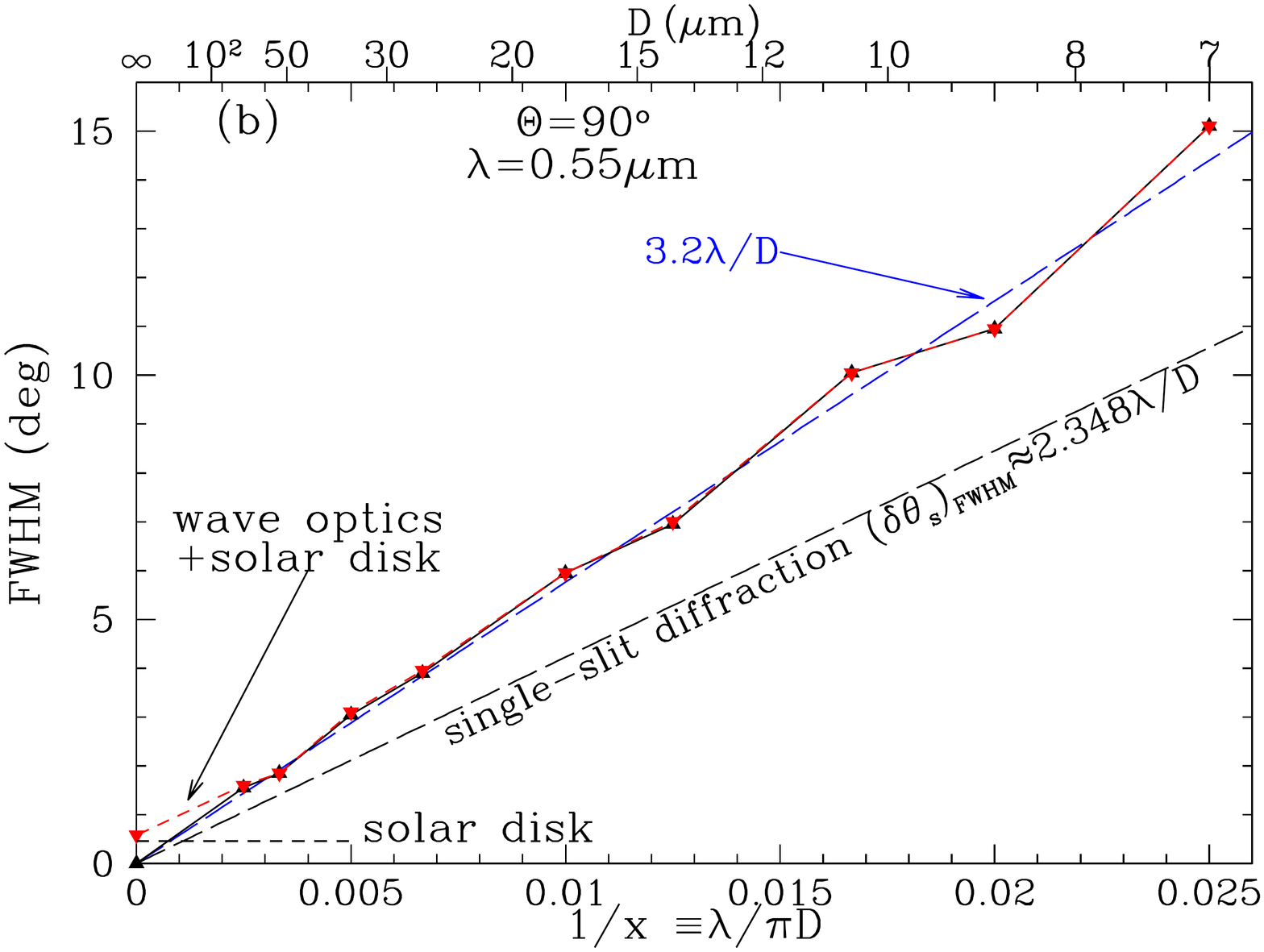}
\caption{\label{fig:particle-sizing} 
         (a)
         The ``power spillover index'' $\Psi$ as a function of $x$
         for $\Theta=90^\circ$.
         $\Psi$ is a measure of scattered power interior to $21.9^\circ$
         (see Eq. \ref{eq:Psi}).
         Line connecting calculated points is only to guide
         the eye.
         (b) Full width at half-maximum (FWHM) for the scattering peak
         near 22$^\circ$, as a function of $1/x=\lambda/\pi D$.
         Line connecting calculated points is only to guide the eye.
         }
\end{figure}

To demonstrate that these effects are not confined to the special
case of $\Theta=90^\circ$, 
Fig.~\ref{fig:S_11_Theta=60} shows results of both DDA and GO
calculations for
incidence angle $\Theta=60^\circ$.
The DDA calculations take the birefringence into account exactly,
but the ray-tracing code 
\cite{macke1996single}
is for isotropic material.
To approximate the results for weakly anisotropic H$_2$O ice,
we take
a weighted sum of separate GO calculations for
refractive indices $n_o$ and $n_e$, with weights $0.5(1+\cos^2\Theta)$
and 
$0.5\sin^2\Theta$ 
for the ordinary ($\bE\perp \bc$)
and extraordinary ($\bE\parallel \bc$) cases.
This treatment is correct for $\Theta=90^\circ$, 
but only an approximation when $\Theta<90^\circ$.

For $\Theta=60^\circ$ incidence, the GO caustics are at 
$\theta_{\rm s,min}=24.90^\circ$ and
$25.02^\circ$ for $n_o$ and $n_e$
(see Eq.\ \ref{eq:thetas_min}).
The DDA results again show convergence to the GO
solution as $x\rightarrow\infty$.
As for $\Theta=90^\circ$, the halo is clearly visible for $x=100$, 
although appreciably broadened
\cite{mishchenko1999big},
with the halo
increasing in peak
intensity and in sharpness as the size parameter increases.

When the sun is not at the zenith, the angle between the incident
light and the {\it c-}axis of the horizontally-oriented crystals will
range from $\Theta=90^\circ$ to $90^\circ-\theta_z$, where $\theta_z$ is
the zenith angle.
Thus the present results, for only $\Theta=90^\circ$ and $\Theta=60^\circ$,
can only be compared to the $22^\circ$ halo for the sun at the zenith
($\theta_z=0$).

\subsection{Particle sizing: power spillover index}

For finite size crystals, ray optics ceases to be a good approximation,
as is evident from comparison of the GO and wave optics results in
Figs. \ref{fig:S_11_Theta=90} and \ref{fig:S_11_Theta=60}.
For $x\leq 400$,
an appreciable amount of scattered light spills into the ``forbidden'' 
(so-called ``shadow'')
region with $\theta_s<\theta_{\rm s,min}$.
The power scattered into this forbidden zone can be used to constrain 
the diameter of
the ice crystals doing the scattering.

We define a quantitative ``power spillover index''
\beq \label{eq:Psi}
\Psi = 
\frac{\int_{19^\circ}^{21.9^\circ} 
(dC_{\rm sca}/d\theta_s) d\theta_s}
{\int_{19^\circ}^{25^\circ} 
(dC_{\rm sca}/d\theta_s) d\theta_s}
~~~.
\eeq
The upper limit $21.9^\circ$ adopted for the integral in the numerator
in Equation (\ref{eq:Psi}) is chosen to be just interior to the
inner edge of the halo for the ordinary polarization, for $\Theta=90^\circ$.
In the GO limit, the only power interior to $21.9^\circ$ comes from
processes other than refraction through faces $j$ and $j+2$, 
thus $\Psi$ is small: $\Psi=0.0060$.

Figure~\ref{fig:particle-sizing}(a) shows the power spillover index $\Psi$
as a function of $x\equiv \pi D/\lambda$
for $\Theta=90^\circ$.
For $x\approx100$ ($D\approx18\mu$m), we see that
$\Psi\approx 0.4$ -- much larger than the value for GO.
It is evident that measurement of the scattered power
interior to the GO caustic at $\theta_{\rm s,min}=22^\circ$ 
provides information about the particle size.
The diagnostic is most straightforward if the light source is at the
zenith, so that all of the scattering is for $\Theta=90^\circ$.
If the light source is at zenith angle $\theta_z>0$, 
then one must take an
appropriate average over $\Theta$ values
between $90^\circ$ and $90^\circ-\theta_z$.

If the sun (or full moon) is the source of illumination, then the
differential scattering cross section
must be convolved with the $0.53^\circ$
diameter uniform brightness disk.
This solar convolution increases the GO value of $\Psi$ from $0.006$
to $0.058$.
For $x\leq400$ the solar convolution has only a minimal effect on the
wave optics results for $\Psi$ [see Fig.~\ref{fig:particle-sizing}(a)].

Of course, interpretation of atmospheric scattering must
recognize that particles
other than long ice columns may also be contributing
to scattering at $\theta_s<\theta_{\rm s,min}$, so that 
a metric like $\Psi$ provides only a lower limit on the diameter
of the halo-producing hex columns. 

\subsection{Particle sizing: FWHM}

Another particle-size 
diagnostic is provided by the FWHM of the scattering peak near $22^\circ$.
In the GO limit, this peak is a caustic, 
with $dC_{\rm sca}/d\theta_s\rightarrow\infty$ and
FWHM~$=0$.
For finite $D$, 
the peak $dC_{\rm sca}/d\theta_s$ is finite, and
FWHM~$>0$.

The $22^\circ$ halo is produced by radiation that enters through a side
face $j$ (see Fig.~\ref{fig:hex})
with width perpendicular to the beam 
$w=(D/2)\cos(30^\circ+\zeta_{\rm min}/2)=0.377D$ 
for $\zeta_{\rm min}=22^\circ$.
Diffraction through a slit $w$ has
\beq \label{eq:diff_FWHM}
(\Delta\zeta)_{\rm FWHM}=0.886 (\lambda/w)=2.348 (\lambda/D)
~~~.
\eeq
Figure~\ref{fig:particle-sizing}(b) shows the FWHM for our wave optics
solutions for $\Theta=90^\circ$.
The FWHM of the $22^\circ$ halo is approximately linear in $\lambda/D$,
although $\sim$40\% larger than Eq. (\ref{eq:diff_FWHM}),
presumably because of differential 
refraction, followed by additional
diffraction upon exit through face $j+2$.
Our results
are approximated by
\beq
(\Delta\zeta)_{\rm FWHM} \approx 3.2 (\lambda/D)
~~~.
\eeq
We have also smoothed the results using a $0.53^\circ$
diameter disk [see Fig.~\ref{fig:particle-sizing}(b)]
but the effects are minimal for $x\leq400$.

It should be stressed that the FWHM in Fig.~\ref{fig:particle-sizing}(b)
is for the special case of $\Theta=90^\circ$ (i.e., sun at the zenith, and
horizontally-oriented hex columns).
For the more general case of solar zenith angle $\theta_z>0$, the FWHM must
be recalculated with appropriate averaging over $\Theta$ values between
$90^\circ$ and $90^\circ-\theta_z$.
Integration over a range of $\Theta$ values will increase the FWHM, with
the effects most pronounced for large values of $D$.

\subsection{Polarization}

For $x\rightarrow\infty$,
application 
\cite{truilo2002astrophysical}
of the Fresnel
equations shows that the halo light is linearly polarized.
If birefringence is neglected, for $\Theta=90^\circ$ 
light passing through faces $j$ and $j+2$, 
the fractional polarization at
$\theta_s=\theta_{\rm s,min}$ is
\begin{eqnarray}
\label{eq:fresnel}
P_h &=& 
\frac
{(n\sqrt{3}+\sqrt{4-n^2})^4- (\sqrt{3}+n \sqrt{4-n^2})^4}
{(n\sqrt{3}+\sqrt{4-n^2})^4+ (\sqrt{3}+n \sqrt{4-n^2})^4}
\\ \nonumber
\label{eq:fresnel2}
&=& 0.037 ~{\rm for}~n=1.312.
\end{eqnarray}
directed parallel to the scattering plane
($P>0$),
i.e., perpendicular to the halo arc.
However,
birefringence also contributes to the polarization.
\begin{figure}[htbp]\centering
\includegraphics[angle=270,width=12.5cm,
                 clip=true,trim=0.5cm 0.5cm 0.5cm 0.5cm]
                {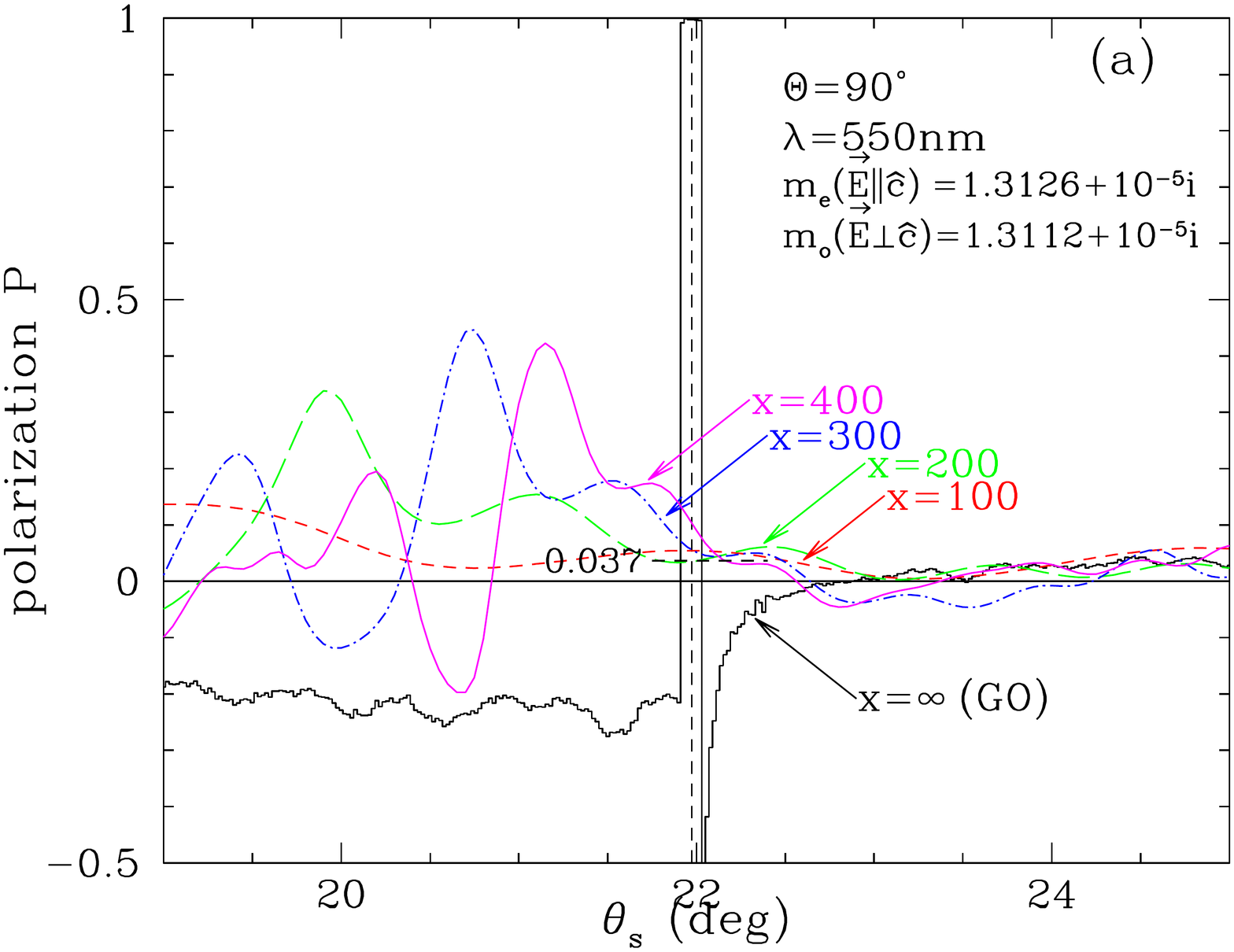}
\includegraphics[angle=270,width=12.5cm,
                 clip=true,trim=0.5cm 0.5cm 0.5cm 0.5cm]
                {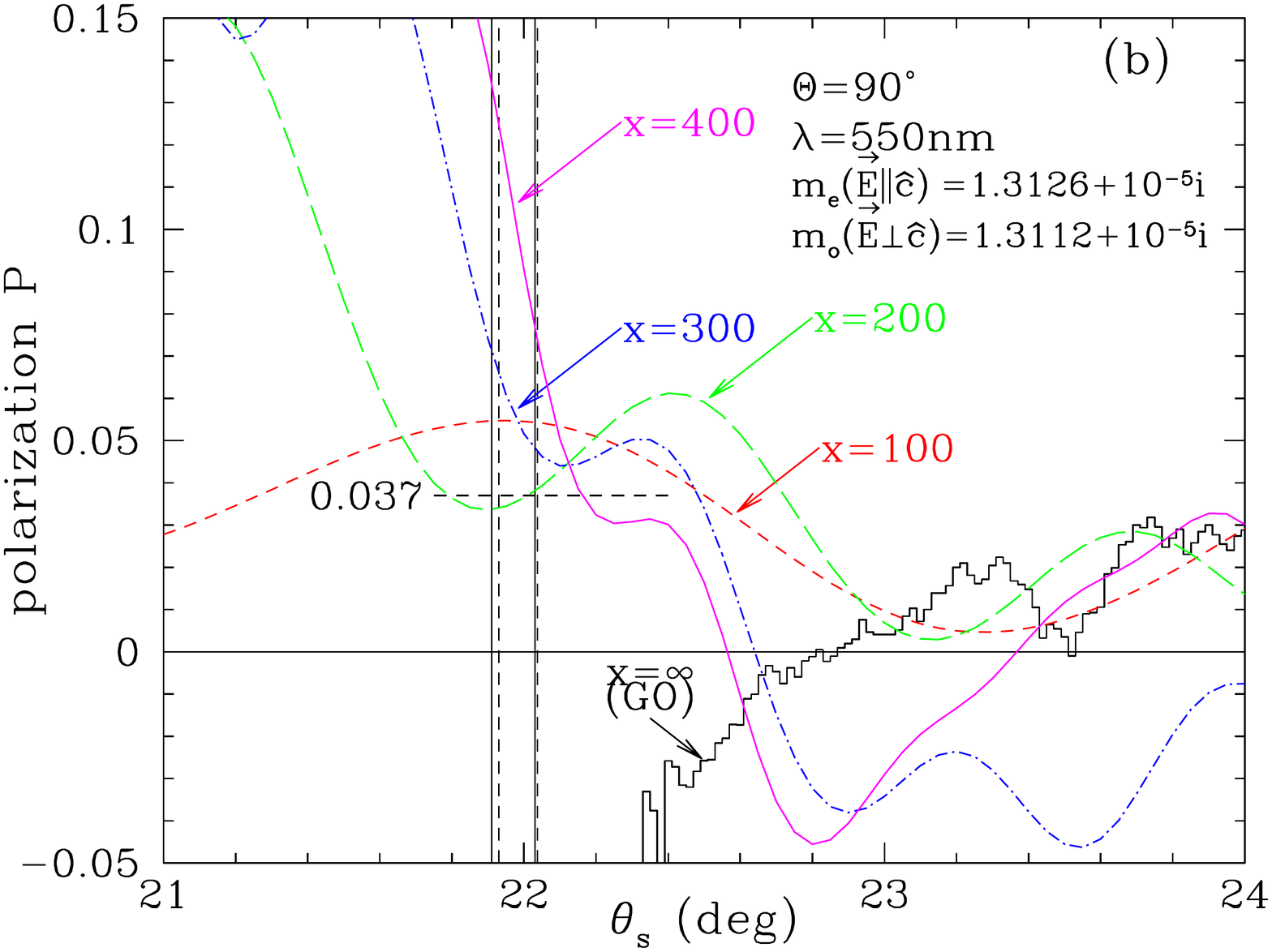}
\caption{\label{fig:pol} 
  Polarization $P=S_{21}/S_{11}$ 
  as a function of scattering angle $\theta_s$
  for hexagonal prism with unpolarized light
  incident perpendicular to the {\it c}-axis ($\Theta=90^\circ$),
  for $x=100,200,300,400$.
  (a) $19^\circ < \theta_s <25^\circ$;
  (b) zoom into $21^\circ <\theta_s < 24^\circ$.
  Also shown (labelled $x=\infty$)
  are GO results, including birefringence.
  Vertical dashed lines show location of halo edge for
  ordinary and extraordinary polarizations.
  Horizontal dashed line shows $P_h=0.037$ expected
  for GO neglecting birefringence (see Eq.~\ref{eq:fresnel}).
  See text for discussion.
  }
\end{figure}

K\"onnen and co-workers 
\cite{konnen1991polarimetry,konnen1994halo}
have discussed the expected birefringence peak in the polarization
for geometric optics.
For $\Theta=90^\circ$: 
(1) For 
$\theta_s<21.93^\circ$, the $22^\circ$ halo is not present,
and $P\approx-0.25$ from weak small-angle
scattering by oblique reflections off the edge faces
(see Fig.~\ref{fig:S_11_Theta=90}).
(2) For $21.930^\circ<\theta_s<22.037^\circ$ the $22^\circ$ halo
is $\sim$100\% polarized with $\bE\perp \bc$, i.e., parallel
to the scattering plane ($P>0$).
(3) Just beyond $22.037^\circ$, both polarizations contribute
to the $22^\circ$ halo, but
the contribution with $\bE\parallel \bc$ will dominate at and just
outside the $22.037^\circ$ caustic, and the net polarization will be
perpendicular to the scattering plane ($P<0$).
(4) At larger angles, $\theta_s>24^\circ$,
the polarization has a value close to
$P_h\approx0.037$ from eq.\ (\ref{eq:fresnel}).
Figure~\ref{fig:pol} shows the polarization over the 20-25$^\circ$
range for $x\rightarrow\infty$
and $\Theta=90^\circ$,
calculated for $x\rightarrow\infty$ using ray-tracing
for random rotations of the target around the {\it c}-axis.

%
K\"onnen and Tinbergen
\cite{konnen1991polarimetry} 
discussed the diffractive
broadening of the $22^\circ$ polarization peak produced by birefringence, 
and showed
how multiwavelength polarization observations 
could be used for particle sizing. 

Figure~\ref{fig:pol} 
shows the polarization obtained from the
full solutions to Maxwell's equations, including birefringence.
Our exact results for $x=100-400$ show pronounced oscillations with
angle, but
have generally positive $P$ in the 20-22$^\circ$ region where the
scattered light is dominated by
``spillover'' of the $22^\circ$ halo into this forbidden region.
For $21.930^\circ<\theta_s<22.037^\circ$ where 
GO
predicts $P\approx1$, the $x=400$ solution
(corresponding to diameter $D\approx70\mu$m)
has $P\approx0.10$ at 550\,nm, 
comparable to $P=0.10$ at 615\,nm and $0.16$ at
435\,nm observed by K\"onnen, Wessels, and Tinbergen 
\cite{konnen2003halo}.
From the progression of $P$ values at $x=200,300,400$, it appears
that $D\approx 90\mu$m ($x\approx500$ for $\lambda=0.55\mu$m) 
would likely reproduce the polarizations
reported by K\"onnen et al.

\section{Summary}
The public-domain code DDSCAT 
\cite{flatau1994discrete}
was used to
calculate light scattering by infinite hexagonal ice prisms for size
parameter of up to $x=400$ in the discrete dipole approximation. 
Birefringence is fully included in the discrete dipole approximation calculations. 
The $22^\circ$ halo like feature is predicted for size parameter as low
as $x=100$, but with an angular structure very different from the
halo calculated using geometric optics.
The halo peak becomes more pronounced and narrower for
increasing size parameters.

The results show that
accurate solutions 
of
Maxwell's equations can be used to relate observed properties of
atmospheric optics displays, including some aspects of halos 
\cite{tape1994atmospheric},
to the particle size.
The power spillover index $\Psi$ [Fig.~\ref{fig:particle-sizing}(a)]
and the FWHM [Fig.~\ref{fig:particle-sizing}(b)] are both measurable
and directly related to the diameter $D$ of the hexagonal columns.
Future research may employ the methodology demonstrated in this
paper to study the effects 
on halo properties
of various factors,
such as porosity
\cite{shcherbakov2013why}
or non-evenness of side edges 
\cite{konnen1994halo};
such target geometries can be treated by DDSCAT.

\section*{Acknowledgments}
BTD was supported in part by NSF grant AST-1008570.
We thank Dr.\ Gunther K{\"o}nnen for several enlightening discussions about
the birefringence of ice and its polarization signature and comments on 
a draft version of this paper,
Dr.\ Andreas Macke for making available his ray-tracing program 
and for helpful comments,
and
Dr.\ Anatoli Borovoi and Dr.\ Andr\'as Barta for providing valuable
understanding of GO scattering by
hexagonal crystals at the $120^\circ$ parhelion.


\end{document}